# Self-organized Nano-lens Arrays by Intensified Dewetting of Electron Beam Modified Polymer Thin-films

*Ankur Verma and Ashutosh Sharma\**

Department of Chemical Engineering & DST Unit on Nanosciences
Indian Institute of Technology Kanpur, Kanpur, UP, 208016 (India)
E-mail: ashutos@iitk.ac.in; Fax: +91-512-259 0104; Tel: +91-512-259 7026

Sub-100 nm polymeric spherical plano-convex nano-lens arrays are fabricated using short electron beam exposures to selectively modify the ultrathin (< 30 nm) polymer films, followed by their intensified self-organized dewetting under an aqueous-organic mixture. A short exposure to e-beam locally modifies the polymer chains to effectively change the viscosity of the film in small domains, thus bringing in the dynamical dewetting contrast in the film that produces aligned and ordered dewetted nanostructures. Both negative and positive e-beam tone polymers are thus used to produce array of nano-lenses. The intensified self-organized dewetting under a water-organic solvent mixture overcomes the limitations on surface tension and dewetting force and thus facilitates the formation of sub-100 nm diameter polymer nanolenses of tunable curvature. By varying the extent of e-beam exposure, various configurations from isolated to connected nano-lens arrays can be fabricated.

**Introduction**

Self-organized pattern generation by dewetting in the ultrathin (< 50 nm) polymer films offers promising routes of fabricating polymeric meso- and nanostructures for potential applications in micro- and nano-optics systems,[1–3] memory devices,[4] polymer electronics[5] and membranes.[6] There are however three major limitations on the pattern formation by self-organization based techniques: (1) Pattern length scale and feature sizes are in micrometers, usually in tens of micrometers owing to a weak destabilizing van der Waals force and a strong stabilizing surface tension force;[7–11] (2) Pattern alignment or ordering requires fabrication of physico-chemically patterned substrates for controlled dewetting;[12–16] and (3) The pattern formed by dewetting in air has small aspect ratios owing to small contact angles of polymer droplets. We report here a novel technique for versatile nano-patterning that removes these limitations by a combination of low dose electron beam lithography (EBL) and self-organized, room temperature dewetting of ultrathin films under a mixture of water and polar organic solvents. Low-dose EBL is employed to first create fine domains of altered polymer viscosity without any loss or complete cross-linking, after which dewetting under an aqueous-organic mix reduces the instability length scale by one to two orders so as to allow selective dewetting of lower viscosity domains in highly confined spaces defined by the EBL pattern. We demonstrate the application of this method for the fabrication of nano-lens arrays where self-organized dewetting can control the lens diameter, curvature and aspect ratio by varying the film thickness and its dewetting time. The latter can tune the contact-angle of the lens in the range of ~30–140°.

Dewetting of a thin polymer film on a nonwettable substrate occurs when it is heated above its glass transition temperature ($T_g$) or its $T_g$ is brought down below room temperature by exposure to solvent vapor. Self-organized dewetting of ultrathin (< 100 nm) liquid polymer films occurs by the attractive inter-surface interactions such as the van der Waals, which make the film unstable resulting in the formation of holes that grow and coalesce to eventually form droplets. The wavelength of long-wave instability is given by:

$$\lambda = [-8\pi^2\gamma/(\partial\phi/\partial h)]^{0.5} \qquad (1)$$

Where $\gamma$ is the interfacial tension, $h$ is the film thickness and $\phi$ is the destabilizing potential (~$h^{-3}$ for van der Waals and ~$h^{-2}$ for electrostatic interactions).[7–10] In air, high energy penalty associated with the formation of new surface at small scales limits the size of these structures to few micrometers even for the films as thin as 10 nm.[7–10] Thus, a significant reduction in the interfacial tension and an intensification of the destabilizing potential are required to achieve much smaller, ~ 100 nm features. Recently, we showed that dewetting under an optimal mixture of water and solvents accomplishes these tasks thus reducing the instability length scales by one to two orders of magnitudes.[3] The solvent molecules diffuse in the polymer chains and bringing down the $T_g$ below the room temperature, but a concurrent dissolution of the polymer in the mixture is inhibited because of water, which is a nonsolvent, being the majority phase. The resulting sub-micrometer droplets could be used as nano-lenses for high resolution optical imaging.[3] The mechanism of instability under the water-organic mixture was also explored recently and found to be about two orders of magnitude reduction in the interfacial tension together with a much stronger destabilizing potential, which is possibly of electrostatic origin.[11] However, a major limitation of the method is that dewetting on a flat homogeneous surface produces randomly distributed droplets. Several strategies have been proposed for aligning the droplets by controlled dewetting on physically or chemically patterned substrates or by combining dewetting with other lithography techniques.[12–16] However, coating a uniform thickness film on physicochemically patterned substrates is often difficult, especially coating of very thin (< 20 nm) films that are unstable. Further, fabrication of a hard (e.g., silicon) template dedicated to dewetting of a single film is an expensive process. Finally, heterogeneities of the template cannot be selectively obliterated after patterning of the polymer. Thus, although template mediated dewetting can give ordered polymer arrays as a proof-of-concept; it is not especially attractive for large-scale manufacturing. It may also be noted that template-mediated dewetting has been demonstrated largely for the feature sizes in the micrometer and tens of micrometer range,[12–16] with a single exception of one study with ~ 100 nm feature sizes.[3,11] Further, it is also important to note that polymeric features with smooth circular shapes, such as nano-lenses and their ordered arrays cannot be directly and easily fabricated with the top-down techniques such as EBL.

Here we present a relatively quick and simple method for overcoming the problems of feature size, aspect ratio, slow kinetics and substrate-induced templating by a combination of electron beam (e-beam) exposure and self-organized dewetting in ultrathin (< 30 nm) polymer films. The basic pattern in the polymer film is first defined by creating significant spatial viscosity differentials by use of a low exposure e-beam tool. Selective and sequential dewetting of highly confined regions under an optimal water-organic mixture then rapidly produces the required pattern by progressive dewetting of increasing viscosity nano-domains. The key to the success of this technique is in the orders of magnitude reduction in the instability length scale under the water-organic mixture compared to air, which allows dewetting to proceed under severe confinements imposed by the e-beam pattern.

**Results and discussion**

Several polymers are sensitive to the e-beam radiation and get modified differently with various degrees of exposure. Polymethylmethacrylate (PMMA) is a well-known positive tone e-beam resist where e-beam exposure reduces the molecular weight by chain breaking and thus produces a low viscosity domain upon melting or exposure to a good solvent. In contrast, polystyrene (PS) is known to behave as a negative tone e-beam resist at low e-beam exposures, which produce higher chain entanglement or weak cross-linking of the polymer chains,[17] thus increasing its viscosity. E-beam exposure has also been shown to be effective for localized cross-linking of block polymers containing polystyrene.[18] Therefore, e-beam exposure at low doses can produce a significant viscosity contrast in the exposed and unexposed areas. Dewetting is initiated in the lower viscosity domains, i.e., unexposed areas of a PS film and exposed areas of a PMMA film. The growth rate of instability and dewetting are significantly higher in these regions. Change in viscosity affects the kinetics of dewetting while length scale of dewetting remains unaffected by this. This is in accordance to the theoretical predictions of spinodal instability where growth rate is inversely proportional to the viscosity and the wavelength of instability is independent of viscosity.[19] Ordered pattern formation is observed when the dimensions of the uncrosslinked regions produced by the e-beam exposure (or its lack) are nearly commensurate with the instability wavelength of homogeneous film on flat substrate. **Figure 1** shows the schematic diagram of the process where a thin film with parallel lines exposed by the e-beam is dewetted.

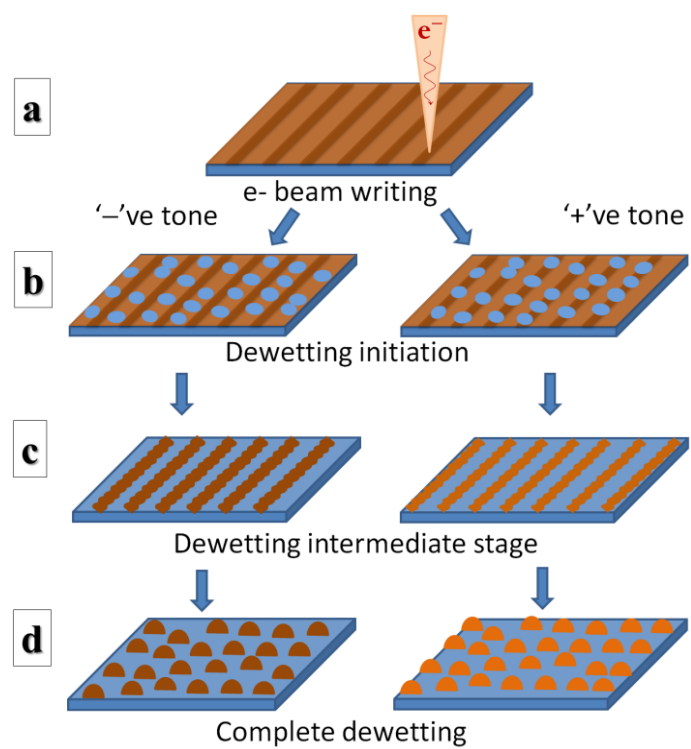

**Figure 1.** Schematic diagram of e-beam modified enhanced dewetting of ultrathin polymer film with negative or positive tone for e-beam. (a) e-beam exposed simple patterns on the polymer film coated on silicon. (b) Dewetting starts after immersion of the film in the water-solvent mixture with the formation of isolated holes that form in the unexposed areas of the film in a negative tone film and in the exposed areas of the film in a positive tone film. (c) These holes grow in size and coalesce, making polymer ribbons. (d) The polymer ribbons eventually break up to form aligned droplets.

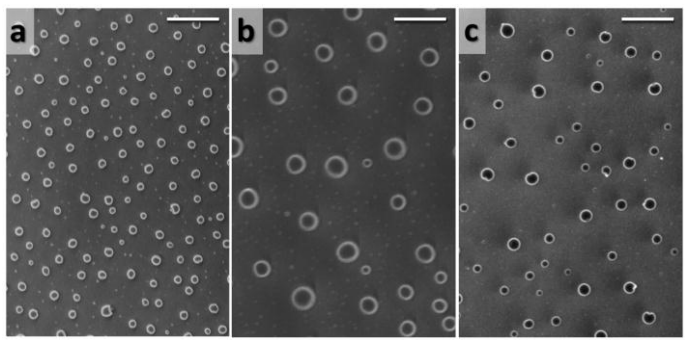

**Figure 2.** FESEM image of dewetted structures of: (a) 6.5 nm thick PS film, (b) 12 nm thick PS film and (c) 9 nm thick PMMA film. (Scale bar: 500 nm)

Dewetting of unexposed homogeneous films is investigated to obtain the droplet size and the spinodal wavelength of dewetting corresponding to a particular film thickness. Dewetting produces smooth structures of tunable curvature (e.g., spherical lenses) which cannot be fabricated by direct e-beam writing. **Figure 2** shows the morphology of dewetted PS films with initial thickness 6.5 and 12 nm and a PMMA film with initial thicknesses 9 nm. The droplet diameters ($d_D$) for the 6.5 nm and 12 nm PS films are 88±17 nm, and 191 ± 36 nm, respectively. In case of the 9 nm thick PMMA film, droplet diameter is 116±26 nm. The spinodal wavelengths ($\lambda_D$) for these films are found to be 235±39, 468±47 and 405±52 nm respectively. It takes less than 30 seconds for dewetting to complete in all the cases. An interesting observation are the extremely tiny droplets in the background which form during the growth of instability. The size of these droplets is about an order of magnitude smaller than the larger drops, thus creating a bi-modal distribution. These smaller secondary droplets form only when the film thickness is very small, less than about 20 nm, but not in the thicker films. The smaller droplets form during the hole-growth owing to an increased effects of slippage and the cross sectional curvature of the hole-rims, both of which promote the formation of unstable fingers that disintegrate into droplets.[20] Here, we focus on the relatively larger droplets for the statistical analysis of droplets size and distribution. In the previous works[3,11] we have published complete data set for the droplet size and wavelength as a function of film thickness for PS and based on the best fit, mean droplet size ($d_D$) and mean wavelength ($\lambda_D$) can be related to initial film thickness ($h$) as:[11]

$$\lambda_D = 6.9 \times 10^6 \cdot h^{1.5} \tag{2}$$

$$d_D = 3.09 \times 10^3 \cdot h^{1.26} \tag{3}$$

The standard deviation is within 30% of the mean value in both the cases. These relations can be used to select appropriate thickness of the polymer film to obtain lenses of required size. The e-beam patterning process described here can also be used to fabricate bigger lenses up to the size of few tens of micrometers.

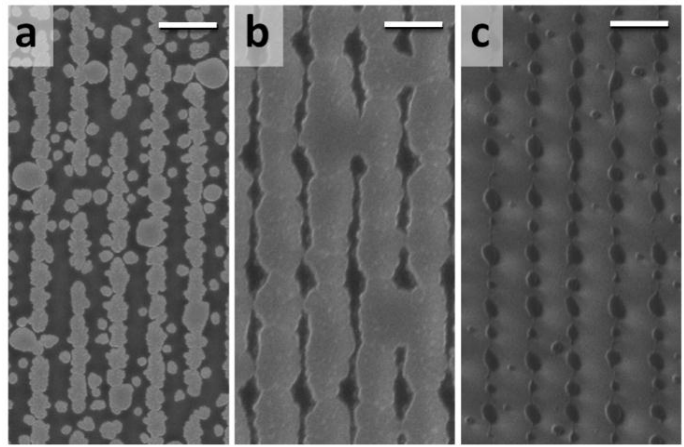

**Figure 3.** Intermediate dewetting stages of a line-patterned e-beam exposed PS thin film of thickness 12 nm after: (a) 2 seconds, (b) 5 seconds and (c) 30 seconds. Dark and light regions are polystyrene and silicon wafer. (Scale bar: 500 nm)

- 7 -

As discussed earlier, unconfined dewetting on a uniform substrate engenders randomly distributed polymer droplets of relatively wide size distribution. Alignment of droplets in the confinement produced by the e-beam exposure is discussed next. Two types of e-beam patterns are drawn on the film: parallel lines and square array of dots. Pitch of the patterns ($\lambda_P$) is chosen to closely match with the dewetting wavelength ($\lambda_D$) of homogeneous unexposed films. The latter can be tuned by changing the film thickness. A very low e-beam dose of 1–10 $\mu C\ cm^{-2}$ was used to expose the polymer films. This is in contrast to the typical values of e-beam doses used in the conventional EBL that range between 50–1000 $\mu C\ cm^{-2}$. Thus, although EBL employed here is still a serial process, it is found to be fairly rapid owing to low doses. A single beam could produce a writing area coverage of about 0.1 $mm^2$ per second (about a million features per second). Films exposed to the e-beam pattern are immersed in to a dewetting solution for 5 – 10 minutes to ensure complete dewetting. Instabilities in the unexposed region of PS film and exposed region of PMMA film grow faster owing to the lack of cross-linking (lower viscosity) in these regions. Various stages of dewetting are shown in **Figure 3** for a PS film. Dewetting starts from the nucleation of holes in the unexposed areas between the exposed lines. The growth of holes displaces the polymer along the exposed lines. At a later time, the polymer line of higher viscosity in the exposed region also dewets to form droplets that are aligned along the line of e-beam exposure.

On a 6.5 nm thick PS film, best 1-D alignment of dewetted droplets is found when the line spacing ($\lambda_P$) is at 220 nm (**Figure 4**a). Spacing of droplets along a line is found to be 242±12 nm. Both of these length scales are in conformity with the dewetting wavelength of a 6.5 nm homogeneous film: ~235±39 nm. The droplet size is 72±15 nm, which is also close to the droplet size produce by dewetting of a homogeneous film, 88±17 nm. Aligned droplets are observed in the 12 nm thick PS film when $\lambda_P$ is 440 nm (Figure 4b), which produces inter-droplet spacing along a line to be also larger at 428±24 nm. This agrees with the dewetting wavelength of the 12 nm homogeneous film, 468±47. The size of the droplets is 158±21 nm, as opposed to 191±36 nm obtained for homogeneous random dewetting. Similar patterns are formed in a 9 nm thick PMMA film and are shown in Figure 4c at e-beam line spacing of 430 nm. Here, aligned droplets of 95±18 nm diameter are formed with an inter-droplet spacing of 283±34 nm. Apparently, 1-D confinement brings in a small further reduction in the droplet size and improves the monodispersity of droplets.

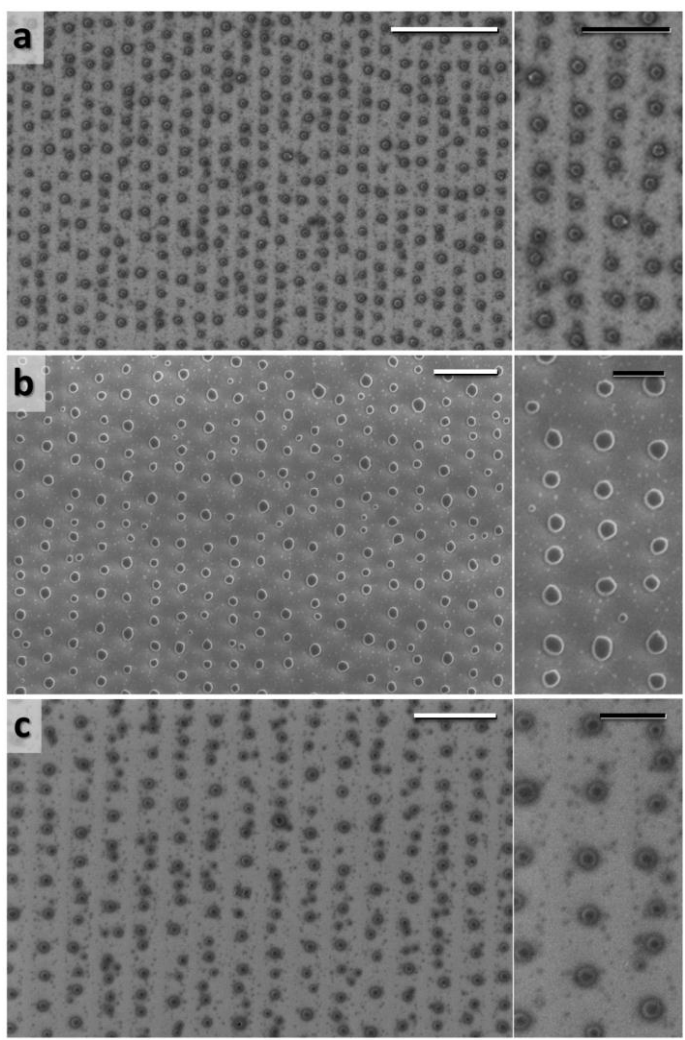

**Figure 4.** 1-D aligned droplets obtained after the dewetting of e-beam modified polymer thin films. (a) 6.5 nm thick PS film resulting in average number density of nanolenses as $18.8 \times 10^6$ per $mm^2$, (b) 12 nm thick PS film giving average number density of nanolenses as $5.3 \times 10^6$ per $mm^2$ and (c) 9 nm thick PMMA film resulting in average number density of nanolenses as $8.2 \times 10^6$ per $mm^2$. (Scale bar: white is 1 µm and black is 400 nm)

To appreciate the importance of the intensification of instability in the success of this technique, a similar dewetting experiment is carried by exposure to a solvent vapor rich air that can induce dewetting by reducing the $T_g$ below the room temperature. The e-beam exposed regions do not dewet at all in air (see **Figure 5**a) unlike the similar case of dewetting under water-organic mixture (Figure 4b). However, when a sparser pattern with a line spacing of 1 μm is exposed by the e-beam on a similar 12 nm thick PS film, incomplete dewetting is observed with much larger irregularly shaped droplets with 2–3 μm of inter-droplet spacing. Moreover, these droplets have very small aspect ratios owing to their very low contact angles (< 10°).

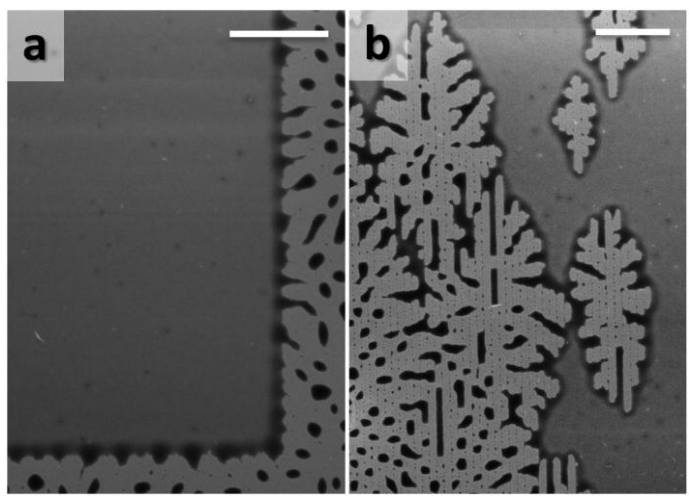

**Figure 5.** Dewetting of an e-beam modified 12 nm thick PS film in solvent vapor rich air. (a) No dewetting occurs in the e-beam exposed part even after 3 hours because of the e-beam confinement length scale being much shorter than the instability length scale in air; (b) Incomplete and disordered dewetting is observed on sparser e-beam patterns with inter-line spacing of 1 μm. (Scale bar: 10 μm)

While 1-D confinement brings some degree of order in the dewetted pattern, better results are obtained by 2-D confinement as shown below by employing a square array of single pixel dots exposed by e-beam. Regularly aligned droplets are formed by dewetting of 6.5 nm PS film when the periodicity of dots, $\lambda_P$ is 225 nm (see **Figure 6**a). The droplet size in this case is found to be 81±12 nm. A well-ordered array is obtained at $\lambda_P$ of 480 nm for the 12 nm thick film (see Figure 6b). The droplet size in this case is found to be 148±21 nm. Thus, ordered dewetting engendered by confinement produces similar pattern length scales and feature sizes as in the case of random dewetting of a corresponding homogeneous film.

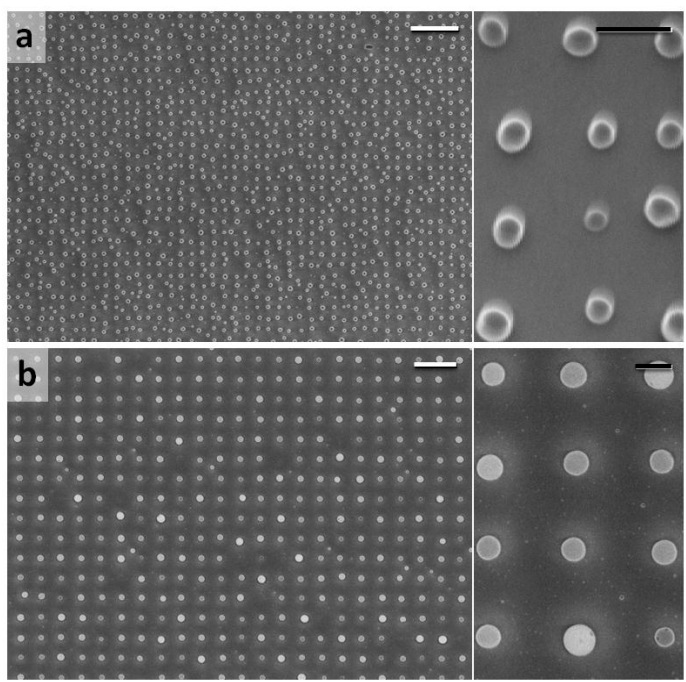

**Figure 6.** 2-D aligned droplets obtained after the dewetting of e-beam modified PS thin film. (a) 6.5 nm thick PS film giving average number density of nanolenses as $19.8 \times 10^6$ per mm$^2$, (b) 12 nm thick PS film resulting in average number density of nanolenses as $4.3 \times 10^6$ per mm$^2$. (Scale bar: white is 1 μm and black is 200 nm)

Next, we discuss the effect of higher e-beam exposures on the dewetted pattern morphology using line exposures. As the e-beam exposure time (or its dose) is increased, qualitatively different morphology patterns can be fabricated and the line pattern morphology approaches towards the conventional EBL. As shown in **Figure 7**a, when the dwell time is doubled (as compared to the case which produced linear array of isolated droplets) the linear array of droplets are connected with a thin line owing to the very high viscosity of polymer in the exposed area which inhibits its dewetting. With a further increase in the e-beam dose, the line width also increases (Figure 7b). On a sparser line pattern, a similar morphology of droplets connected with a line is seen (Figure 7c). When very high dose of e-beam is used on a similar film, the line pattern that is similar to the one formed by the conventional EBL of negative e-beam resist is obtained except for the polymer in between the lines which dewets to form aligned droplets in-between the lines (Figure 7d). The connected droplets shown in Figure 7a-c were found to be very advantageous in holding the nanolenses in a mesh, thus facilitating their clean detachment from the substrate and in transferring them on a device or a test sample.

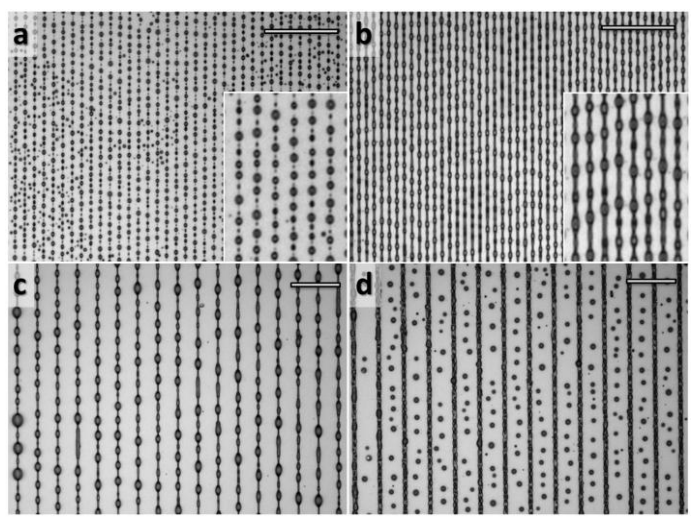

**Figure 7.** Different morphologies of self organized structures on PS thin films with varying e-beam exposure. (a) Linear array of nano-lenses connected with a very thin line of polymer ($h$ =20 nm, $t_d$ = 1000 ns). (b) Higher e-beam exposure yields nanolenses connected with thicker lines ($h$ =30 nm, $t_d$ = 1500 ns). (c) Sparser line patterns (8 μm) produce similar morphologies ($h$ =30 nm, $t_d$ = 2000 ns). (d) Higher exposure along sparser lines (12 μm) yields line structures similar to the conventional EBL together with the dewetted nanolenses in the intervening areas ($h$ =30 nm, $t_d$ = 4000 ns). Here $h$ is film thickness and $t_d$ is dwell time (e-beam exposure time on a single point) (Scale bar: 20 μm)

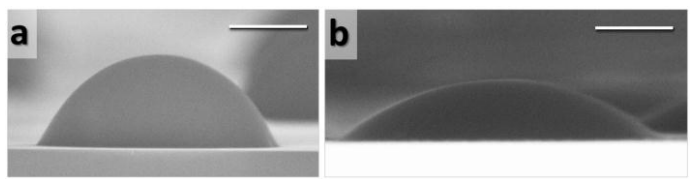

**Figure 8.** Transverse FESEM image of polystyrene nano-lens of tunable shape. (a) 155 nm lens fabricated by enhanced dewetting for 10 minutes. (b) After annealing at 150 °C for 30 minutes in air it spreads out to form 208 nm lens with smaller curvature. (Scale bar: 50 nm)

The contact angle of dewetted polymer droplets increases gradually after their formation. For example, PS droplets attain contact angle of about 60–70° after ~10 minutes of dewetting as shown in **Figure 8**a. When these structures are annealed in air above $T_g$ (~150 °C) for about 30 minutes, the contact angle reduces to 35–40° (Figure 8b), with a concurrent increase in the diameter of the droplet from 155 nm to 208 nm. The contact angle can also be increased further to a maximum of ~ 140° after about 1 hour of immersion in the water-organic mixture. Thus, the size, curvature and the aspect-ratio of the droplets can be tuned over a wide range making them potential candidates for use as nano-lenses in photonics and optics, including near-field imaging.[2]

**Experimental**

Ultrathin (6–30 nm) films of PS (Mw: 280 kg mol$^{-1}$) and PMMA (Mw: 120 kg mol$^{-1}$) were spin coated on thoroughly cleaned silicon wafers with native oxide layer by using polymer solutions in toluene. Films were then annealed at 60°C for 12 hours in a vacuum oven. Film thickness was measured using nulling ellipsometer (Nanofilm, EP3-SE). Xenos XeDRAW e-beam lithography (EBL) system was used to draw simple single pixel (~10 nm) patterns in the form of lines or dots on the film. Very low e-beam dose (1–10 μC cm$^{-2}$) was used for the exposure with dwell time ($t_d$) 100–1000 ns at 15 kV e-beam accelerating voltage except in the high e-beam dose experiments as shown in Figure 8. E-beam modified films were immersed in a mixture of water, acetone and butanone (15:3:7) for few minutes (< 10 min) and then dried in hot air. Imaging of the fabricated structures was done using field emission scanning electron microscope (FESEM, Zeiss Supra 40VP) and optical microscope (Zeiss Axio observer Z1).

**Conclusions**

In conclusion, the intensified self-organized dewetting of the e-beam modified polymer films under an aqueous-organic mixture provides a simple and rapid method of fabrication for sub-100 nm polymer lens arrays. These structures cannot otherwise be made by a direct write method owing to their smooth circular contours. Both negative (PS) and positive (PMMA) e-beam tone polymers are shown to work equally well. An array of 80 nm PS nanolenses over a large area (~mm$^2$) could thus be easily fabricated in less than 10 minutes of fabrication time. A wide range of nano-lens configurations from isolated to connected arrays can be fabricated by varying e-bam dose, initial film thickness and the dewetting time. The method works by creating a differential contrast of viscosity and a sequential dewetting of fine structures starting from the lower viscosity domains. Compared to dewetting in air, dewetting under the water-organic mixture greatly enhances the surface instability by overcoming the surface tension limitation and by intensifying the destabilizing force. As a result, a much faster dynamics of pattern formation is achieved together with more than one order reduction in the feature size. The technique is both significantly faster than the electron beam lithography and more importantly, can be used to shape polymeric nano-domains with smooth contours. These polymer nanolenses can also be used as the active centers in nanophotonics[3,4] and can have potential applications in fabricating nano-arrays of functional materials such as embedded catalyst nanoparticles.[21]

**Acknowledgements**

This work is supported by a Department of Science and Technology, India by an IRHPA grant and Unit of Nanosciences at IIT Kanpur.